\documentstyle [12pt]{article}
\def\o{\over}
\def\A{\rightarrow}

\def\a{\alpha}
\def\b{\beta}

\def\p{\pi}
\def\s{\sigma}
\def\th{\theta}
\def\vp{\varphi}

\def\bar{\overline}

\def\t{\tilde}
\def\l{\lambda}
\def\G{{\rm GeV}}
\begin{document}
\baselineskip=24pt
\setcounter{page}{1}
\thispagestyle{empty}
\topskip 2 cm
%%%%%%%%%%%%%%%%%%%%%%%%%%
\begin{flushright}
\begin{tabular}{c c}
& {\normalsize AUE-95/02}\\
& {\normalsize April 1995}
\end{tabular}
\end{flushright}
%%%%%%%%%%%%%%%%%%%%%%%%%
\vspace{0.01cm}
\centerline{\large\bf Explicit $CP$ Violation of the Higgs Sector in the}
\centerline{\large\bf Next-to-Minimal Supersymmetric Standard Model}
\vskip 1 cm
\centerline{{\bf Masahisa MATSUDA}
\footnote{E-mail:\ masa@auephyas.aichi-edu.ac.jp}}
\vskip 0.3 cm
\centerline{\it{Department of Physics and Astronomy, Aichi
University of Education}}
\centerline{\it Kariya, Aichi 448, JAPAN}
\vskip 0.3 cm
\centerline{and}
\vskip 0.3 cm
\centerline {{\bf Morimitsu  TANIMOTO}
\footnote{E-mail:\ tanimoto@edserv.ed.ehime-u.ac.jp}}
\vskip 0.3 cm
\centerline{
\it{Science Education Laboratory, Ehime University, Matsuyama 790, JAPAN}}
\vskip 1.0 cm
\centerline{\bf ABSTRACT}
 \vskip 0.5 cm
 We study the explicit $CP$ violation of the Higgs sector in the minimal
supersymmetric standard model with a gauge singlet Higgs field.
The magnitude of $CP$ violation is discussed in the limiting cases of
$x\gg v_1,\ v_2$ and $x\ll v_1,\ v_2$, where $x$ and  $v_{1,2}$ denote
VEV of singlet and doublet Higgs scalars, respectively.
Our numerical predictions of the electric dipole moments of electron and
neutron lie around the present experimental upper limits.
It is found that the large $CP$ violation effect reduces the magnitude
of the lightest Higgs boson mass  in the order of a few ten GeV.
\newpage
%%%%%%%%%%%%%%%%%%%%%%%%%%%%%%%%%%%%%%%%%%%%%%%%%%%%%%%%%%%%%%%%%%%%%%%%%%%%%%%
%%%%%%%%%%%%%%%%%%%%%%%%%%%%%%%%%%%%%%%%%%%%%%%%%%%%%%%%%%%%%%%%%%%%%%%%%%%%%%
\topskip 1 cm
\noindent{\bf 1. Introduction}\par
The physics of $CP$ violation has attracted much recent attention in the light
that the $B$-factory will go on line in the near future at KEK and SLAC.
The central subject of the $B$-factory is the test of the standard model(SM),
in which the origin of $CP$ violation is reduced to the phase in the
Kobayashi-Maskawa matrix[1].
However, there has been a general interest in considering other approaches to
$CP$ violation since many alternate sources exist.
The attractive extension of the standard Higgs sector is the two Higgs
doublet model(THDM)[2], yielding both charged and neutral Higgs bosons
as physical states.
The THDM with the soft breaking term of the discrete symmetry demonstrates
explicit or spontaneous $CP$ violation[3,4,5].
On the other hand, the recent measurements of the gauge couplings at $M_Z$
scale  suggest the minimal supersymmetric standard model(MSSM) is a good
candidate beyond the standard model in the standpoint of the  unification[6].
In this model,  $CP$ violation  has been investigated  with the  soft
symmetry breaking terms[7] since there is no $CP$ violating source in the
Higgs sector at the tree level although two Higgs doublets exist[2].
However, the spontaneous $CP$ violation could occur in the neutral Higgs
exchange through a one-loop potential suggested by Maekawa[8] in the MSSM.
Unfortunately, the scenario to violate $CP$ spontaneously by radiative
correction requires a lighter Higgs boson mass[9,10] than its lower limit
obtained at LEP[11].
\par
The spontaneous $CP$ violation in the extended supersymmetric model was
discussed by some authors[12,13,14].
The most challenging approach is to add a gauge singlet Higgs field $N$ to
the MSSM.
This next-to-MSSM(NMSSM) was studied by many authors especially in the
interests of mass spectra of Higgs sectors[15,16].
The detailed analysis of the mass spectra in this model was studied by Ellis
et al.[16], in which  $CP$ violation in the Higgs sector was neglected.
Although there is  a "NO-GO" theorem of the spontaneous $CP$ violation in the
NMSSM[13], the radiative correction may open the way to cause the spontaneous
$CP$ violation  as shown by Babu and Barr[14].
In this scenario, the light Higgs boson is also unavoidable.
On the other hand, additional singlet $N$ could cause  explicit $CP$
violation in the Higgs sector even at tree level.
In this paper, we study the explicit $CP$ violation of the Higgs sector in
the NMSSM  phenomenologically.
The lightest Higgs boson could be heavier than the $Z^0$ boson by including
radiative corrections.
\par
In section 2, the explicit $CP$ violation is studied with the Higgs potential
in general.
In section 3, the magnitude of the explicit  $CP$ violation is discussed in
the special limiting cases of the vacuum expectation values(VEV) of the
singlet Higgs boson $N$.
In section 4, the numerical results are discussed by using the recent
experimental bounds such as masses of Higgs scalars and the electric dipole
moments of neutron and electron.
Section 5 is devoted to summary.
\vskip 1 cm
%%%%%%%%%%%%%%%%%%%%%%%%%%%%%%%%%%%
%%%%%%%%%%%%%%%%%%%%%%%%%%%%%%%%%%%%%%%%%%%%%%%%%%%%%%%%%%%%%%
\noindent{\bf 2. Explicit $CP$ violation in Higgs Potential}
\par
In order to give masses to all the quarks and leptons, and to cancel gauge
anomalies, at least two Higgs doublets $H_1$, $H_2$ are required in a
supersymmetric version of the standard model[7].
Our discussing model is the MSSM to which a gauge singlet Higgs scalar $N$
has been  added with the requirement that the superpotential contains only
cubic terms[15,16] as follows:
\begin{equation}
W=h_U Q u^c H_2+h_D Q d^c H_1+h_E L e^cH_1+\l H_1 H_2 N -{1\o 3}k N^3+
\cdots \ ,
\end{equation}
\noindent where $Q$, $L$, $u^c$, $d^c$ and $e^c$ are usual notations of
quarks and leptons, and the ellipsis stands for possible nonrenormalizable
terms.
The effective scalar potential is given as
\begin{eqnarray}
V_{\rm Higgs}&=& V_F+V_D+V_{soft} \ , \\ V_F&=&|\l|^2[(|H_1|^2+|H_2|^2)
|N|^2+|H_1 H_2|^2]+ |k|^2|N|^4 \nonumber \\
& &-(\l k^*H_1H_2N^{*2}+H.c.)-|\l|^2(H_1^0H_2^0H^{+*}H^{-*}+H.c.) \ , \\
V_D&=&{g^2\o 8}(H_2^\dagger\hat\sigma H_2+ H_1^\dagger\hat\sigma H_1)^2
+{g'^2\o 8}(|H_2|^2-|H_1|^2)^2 \ , \\
V_{soft}&=&m_{H_1}^2|H_1|^2+m_{H_2}^2|H_2|^2+m_N^2|N|^2
-(\l A_\l H_1 H_2 N+H.c.) \nonumber \\
& & -({1\o 3}kA_kN^3+H.c.) \ ,
\end{eqnarray}
\noindent
where
$H_1\equiv (H_1^0, H^-)$, $H_2\equiv (H^+, H_2^0)$,
$H_1 H_2\equiv H_1^0 H_2^0-H^-H^+$ and
$\hat\sigma\equiv (\sigma^1,\sigma^2,\sigma^3)$.
The radiative effect of the top-quark and top-squark is significant for the
mass spectra of the Higgs bosons as pointed out by some authors in the
MSSM[17].
This leading-log radiatively induced potential is given as follows:
\begin{equation}
V_{\rm top}= {3\o 16\pi^2}\left [(h_t^2 |H_2|^2 + M^2_{sq})^2
\ln {{(h_t^2 |H_2|^2+M^2_{sq})\o Q^2}} - h_t^4 |H_2|^4 \ln
{{h_t^2 |H_2|^2 \o Q^2}} \right ] ,
\end{equation}
\noindent
where we have assumed degenerate squarks:
$ M_{\t t_L}=M_{\t t_R}=M_{sq} \gg m_t$.
The potential $V_{\rm top}$ should be added to $V_{\rm Higgs}$ in eq.(2).
\par
In general, $\l$, $k$, $A_\l$ and $A_k$ are complex, however, by redefining
the global phase of the fields $H_2$ and $N$, we can take
\begin{equation}
\l A_\l \geq 0 \ , \qquad\qquad kA_k\geq 0 \ , \end{equation}
\noindent without loss of any generality.
If we allow  $CP$ violation explicitly in the Higgs scalar sector,
$\l k^*$  is a complex.
\par
%%%%%%%%%%%%%%%%%%%%%%%%%%%%%%%%%%%%%%%%%%%%%%
The VEV of the Higgs potential $V_{\rm Higgs}$ is composed of the neutral
sector and the charged sector written as
\begin{equation}
\langle V_{\rm Higgs} \rangle=\langle V_{{\rm neutral}} \rangle
+ \langle V_{{\rm charged}} \rangle \ .
\end{equation}
\noindent
Our discussion is concentrated on the neutral Higgs sector because there is
no $CP$ violation in the charged Higgs sector.
Since the contribution of $V_{\rm top}$ is not important for qualitative
studies of the explicit $CP$ violation,  we discuss the magnitude
of $CP$ violation without $V_{\rm top}$ in sections 2 and 3.
However, $V_{\rm top}$ contributes significantly to the mass spectra of the
Higgs bosons, so we include this effect in the numerical analyses in section 4.
Neglecting  $V_{\rm top}$ for simplicity, we can write
\begin{eqnarray}
\langle V_{{\rm neutral}} \rangle &=&
\l^2(|x|^2|v_1|^2+|x|^2 v_2^2+|v_1|^2 v_2^2)+k^2|x^4|- v_2(\l k^* v_1 x^{*2}
 +\l^* k v_1^* x^2) \nonumber\\
 &&+{g^2+g'^2\o 8}(|v_1^2|-v^2_2)^2+m_{H_1}^2|v_1^2| +m_{H_2}^2v_2^2 +
m_{N}^2|x^2| \nonumber \\
&& -\l A_\l v_2(v_1 x+v_1^*x^*)-{kA_k\o 3}(x^3+x^{*3}) \ , \end{eqnarray}
\noindent where
VEV's of the neutral Higgs scalar fields are defined as follows:
\begin{equation}
v_1\equiv \langle H^0_1\rangle \ , \qquad v_2\equiv \langle H^0_2\rangle \ ,
\qquad x\equiv \langle N\rangle \ .
\end{equation}
\noindent
The VEV's $v_1$ and $x$ are taken to be complex, and $v_2$ is taken to be a
real positive number without loss of generality.
Therefore, $v_1$ and  $x$ are replaced with
\begin{equation}
v_1 \Longrightarrow v_1 e^{i\a} \ , \hskip 3 cm x
\Longrightarrow x e^{i\omega} \ ,
\end{equation}
\noindent
where $v_1$ and $x$ in RHS are redefined to be real positive numbers, and we
give familiar definitions such as
\begin{equation}
\tan\b \equiv {v_2\o v_1} \ , \qquad v^2\equiv v_1^2+v_2^2 \ .
\end{equation}
\noindent We also introduce a phase for $\l k^*$ as follows:
\begin{equation}
\l k^* = \l k e^{i\vp} \ ,
\end{equation}
\noindent where $\l$ and $k$ in RHS are redefined as positive real numbers.
One can use the minimization conditions of $V_{{\rm neutral}}$
to re-express the soft supersymmetric breaking masses $m_{H_1}^2$,
$m_{H_2}^2$, $m_{N}^2$ in terms of the three VEV's and of the remaining
parameters $\l$, $k$, $A_\l$, $A_k$:
\begin{eqnarray}
m_{H_1}^2&=&\l A_\l {v_2 x\o v_1}\cos(\a+\omega)-\l^2(x^2+v_2^2)+\l k
{v_2 x^2\o v_1}\cos(\vp+\a-2\omega) \nonumber \\ & & +{g^2+g'^2\o 4}
(v^2_2-v^2_1) \quad , \nonumber \\
 m_{H_2}^2&=&\l A_\l {v_1 x\o v_2}\cos(\a+\omega)-\l^2(x^2+v_1^2)+\l k
{v_1 x^2\o v_2}\cos(\vp+\a-2\omega) \nonumber \\ & & +{g^2+g'^2\o 4}
(v^2_1-v^2_2) \quad , \nonumber \\
m_{N}^2&=&\l A_\l {v_1 v_2\o x}\cos(\a+\omega)+k A_k x\cos 3\b-
\l^2(v_1^2+v_2^2)-2k^2 x^2 \nonumber \\
& &+2\l k v_1 v_2\cos(\vp+\a-2\omega) \quad  . \end{eqnarray}
The presence of phases $\a$ and $\omega$ allows in principle for the
spontaneous $CP$ violation. This case was discuused numerically by
Babu and Barr[14].
We work in the vacuum of $\a=0$ and $\omega=0$ since we consider the case of
the explicit $CP$ violation in this paper.
\par
%%%%%%%%%%%%%%%%%%%%%%%%%%%%%%%%%%%%%%%%%%%%%%%%%%%%%%%%%%%
Let us study the masses of the Higgs scalars. The physical charged Higgs
fields is given by
\begin{equation}
C^+\equiv \cos\b H^+ +\sin\b H^{-*} \quad ,
\end{equation}
while the orthogonal combination corresponds to an unphysical Goldstone boson.
The physical charged Higgs boson mass is given as follows:
\begin{equation}
m^2_C=m_W^2-\l^2 v^2+\l(A_\l+kx\cos\vp){2x\o \sin2\b} \ , \end{equation}
\noindent where $m_W^2= g^2 v^2/2$.
On the other hand, the neutral Higgs scalar masses are given by $5\times 5$
mass marix.
Decomposing the neutral Higgs fields into their real imaginary components
\begin{equation}
H_1^0\equiv {S_1+i P_1\o \sqrt{2}} \ , \qquad H_2^0\equiv {S_2+i P_2\o
\sqrt{2}} \ , \qquad N \equiv {X+i Y\o \sqrt{2}} \ ,
\end{equation}
\noindent
shifting $H_1^0$, $H_2^0$, $N$ by their expectation values, and expanding
the neutral Higgs scalar part of $V_{\rm Higgs}$, we get the mass matrix of
the neutral Higgs scalars.
After expressing $P_1$ and $P_2$ in terms of the neutral Goldstone boson
$G^0\equiv \cos\b P_1-\sin\b P_2$ and its orthogonal state
$A \equiv \sin\b P_1+\cos\b P_2$, we get $5\times 5$ mass matrix for the Higgs
bosons $A$, $Y$, $S_1$, $S_2$ and $X$ as follows:
%%%%%%%%%%%%%%%%%%%%%%%%%%%%%%%%%%%%%%%%%%%%%
\begin{equation}
M^2_{\rm Higgs} = \left [\matrix{ M^{AY}_{AY} & & M^{AY}_{S_1 S_2 X}\cr
&       & \cr
(M^{AY}_{S_1 S_2 X})^T & & M^{S_1 S_2 X}_{S_1 S_2 X} \cr}\right ]
\ ,
\end{equation}
\noindent where
$M^{AY}_{AY}$, $M^{AY}_{S_1 S_2 X}$ and $M^{S_1 S_2 X}_{S_1 S_2 X}$
are $2\times 2$, $2\times 3$ and $3\times 3$ submatrices, respectively.
The matrix $M^{AY}_{AY}$ is the one for the Higgs pseudoscalars $A$ and $Y$
as follows:
\begin{equation}
M^{AY}_{AY} = \left(\matrix{ {\l xv^2\o v_1 v_2}(A_\l +kx\cos\vp)
& \l v(A_\l - 2kx\cos\vp) \cr  & \cr
\l v(A_\l -2kx\cos\vp)
& {\l v_1 v_2\o x}A_\l +3A_k kx +4\l k v_1v_2 \cos\vp \cr}\right ) \ .
\end{equation}
\noindent
The matrix $M^{S_1 S_2 X}_{S_1 S_2 X}$ is the one
for the Higgs scalars $S_1$, $S_2$ and $X$ as follows:
\begin{equation}
M^{S_1 S_2 X}_{S_1 S_2 X} =
\left(\matrix{ \bar g^2v_1^2
+{\l v_2 x\o v_1} & v_1v_2(2\l^2-\bar g^2) &2\l^2 v_1 x \cr
+{\l v_2 x\o v_1}(A_\l+kx\cos\vp)
&-\l x(A_\l+kx\cos\vp) & -\l v_2(A_\l+2kx\cos\vp)\cr & & \cr
v_1v_2(2\l^2-\bar g^2) & \bar g^2v_2^2 & 2\l^2 v_2 x\cr -\l x(A_\l+kx\cos\vp)
& +{\l v_1 x\o v_2}(A_\l+kx\cos\vp) & -\l v_1(A_\l+2kx\cos\vp) \cr & & \cr
2\l^2 v_1 x & 2\l^2 v_2 x &{\l v_1 v_2\o x}A_\l \cr -\l v_2(A_\l+2kx\cos\vp)
& -\l v_1(A_\l+2kx\cos\vp) & -A_k kx+4k^2x^2 \cr}\right )\ ,
\end{equation}
\noindent where $\bar g^2\equiv (g^2+g'^2)/2$.
The matrix $M^{AY}_{S_1 S_2 X}$ is the mixing terms of the scalar and
pseudoscalar components as follows:
\begin{equation}
M^{AY}_{S_1 S_2 X} = \left(\matrix{ {k\l v_1x^2\o v}\sin\vp & {k\l v_2x^2\o v}
\sin\vp & 2k\l v x\sin\vp \cr
& & \cr
-2k\l v_2 x\sin\vp & -2k\l v_1 x\sin\vp & -2k\l v_1v_2\sin\vp
\cr}\right ) \ .
\end{equation}
\noindent This submatrix is zero if $CP$ is conserved, that is to say,
$\vp=0$. Then, the matrix $M^2_{\rm Higgs}$ in eqs.(18) $\sim$ (21)
is reduced to the one given by Ellis et al.[16].
\par
\vskip 1 cm
%%%%%%%%%%%%%%%%%%%%%%%%%%%%%%%%%%%%%%%%%%%%%%%%%%%%%%%%%%%
\noindent {\bf 3. $CP$ Violation in Special Limiting Cases}\par
 In general, $CP$ symmetry is violated due to the scalar and pseudoscalar
mixing of eq.(21).
Its magnitude depends on the values of the Higgs potential parameters,
especially, x.
Following  analyses of the Higgs mass spectra by Ellis et al.[16], we study
the magnitude of  $CP$ violation in the three special limiting cases:
 (A) $x\gg v_1,\ v_2$ with $\l$ and $k$ fixed, (B) $x\gg v_1,\ v_2$
 with $\l x$ and $kx$  fixed and (C) $x\ll v_1,\ v_2$.
These limits are discussed in the phenomenological standpoint.
\par
%%%%%%%%%%%%%%%%%%%%%%%%%%%%%%%%%%%%%%%%%%%%%%%%%%
\noindent{\bf (A) Limits of $x\gg v_1,\ v_2$($\l$, $k$ fixed)}
\par
In this limit with $A_\l, A_k\simeq O(x)$, the matrix $M^2_{\rm Higgs}$ in
eqs.(18)$\sim$(21) becomes very simple. Remaining only the terms of order
$O(x^2)$, the Higgs scalar $X$ and the Higgs pseudoscalar $Y$ almost decouple
from other Higgs bosons since these mixing terms are at most order $O(x)$.
The masse squares of $X$ and $Y$ bosons are an order of $O(x^2)$ and then,
those mixing  is negligible small.
The effect of $X$ and $Y$ contributes to our result in the order of
$v_1/x$ and $v_2/x$ through the mixings. Therefore, it is enough for $CP$
violation to consider $3\times 3$ submatrix as to $A$, $S_1$ and $S_2$.
Then, the mass matrix is given in the $A-S_1-S_2$ system as follows:
\begin{equation}
M^2_{\rm Higgs} = \left[\matrix{ 2\l x A_\s/\sin 2\b & \l kx^2\cos\b\sin\vp
& \l k x^2\sin\b\sin\vp \cr & & \cr
\l kx^2\cos\b\sin\vp & \bar g^2 v^2\cos^2\b+\l x A_\s\tan\b
& (\l^2-{\bar g^2\o 2})v^2\sin 2\b-\l xA_\s \cr & & \cr
\l k x^2\sin\b\sin\vp & (\l^2-{\bar g^2\o 2})v^2\sin 2\b-\l xA_\s
& \bar g^2 v^2\sin^2\b+\l x A_\s\cot\b \cr}\right ]
\end{equation}
\noindent where $A_\s\equiv A_\l+kx\cos\vp$ is defined conveniently and $A_\s$
is taken to be of $O(x)$.
By rotating this matrix using $U_0$ with
\begin{equation}
U_0 = \left[\matrix{ 1 & 0 & 0 \cr 0 & \cos\b & -\sin\b \cr
0 & \sin\b &\cos\b \cr}\right ] \ ,
\end{equation}
\noindent
we get simple form of the matrix $M'^2_{\rm Higgs}=U_0^T M^2_{\rm Higgs}U_0$
in the new basis of $A-S_1'-S_2'$ as follows:
\begin{equation}
M'^2_{\rm Higgs} = \left[\matrix{ {2\l x A_\s\o\sin 2\b} & \l kx^2\sin\vp & 0
\cr & & \cr
\l kx^2\sin\vp & (\bar g^2 \cos^2 2\b+\l^2 \sin^2 2\b)v^2 & (\l^2-\bar g^2)v^2
\sin 2\b \cos 2\b \cr & & \cr
0 & (\l^2-\bar g^2)v^2\sin 2\b \cos 2\b
&(\bar g^2-\l^2)v^2\sin^2 2\b+{2\l x A_\s\o\sin 2\b} \cr}\right ] \ .
\end{equation}
\noindent In this matrix, the (2-2), (2-3), (3-2) components are very small
because these are  order of $O(v^2)$ but others are $O(x^2)$.
Therefore, the submatrix of $S_1'-S_2'$ system is almost diagonal  one.
Now, we consider only $A-S_1'$ submatrix, which leads to  $CP$ violation,
as follows:
\begin{equation}
M'^2_{\rm Higgs} = \left[\matrix{ {2\l x A_\s\o\sin 2\b} & & \l kx^2\sin\vp
\cr & & \cr
\l kx^2\sin\vp & &(\bar g^2 \cos^2 2\b+\l^2 \sin^2 2\b)v^2 \cr}\right ] \ .
\end{equation}
\noindent
Since this matrix has a hierarchical structure, one should investigate
these mass eigenvalues carefully.
While the pseudoscalar mass is very large as $O(x)$, the scalar mass
 is very small as $O(v)$.
In order to get the condition of positive eigenvalues,
 we take the determinant of this matrix:
\begin{equation}
Det[M'^2_{\rm Higgs}]\geq 0  \ ,
\end{equation}
\noindent
which gives  a constraint  $\l kx^2\sin\vp \leq O(xv)$.
Since $\l$ and $k$ are constants, we get
\begin{equation}
\sin\vp \leq O(v/x) \ ,
\end{equation}
\noindent which means the scalar-pseudoscalar mixing vanishes in the $x\A
\infty$ limit.
Therefore, it is concluded that  $CP$ violation is minor in this limit.
\par
\vskip 0.3 cm
%%%%%%%%%%%%%%%%%%%%%%%%%%%%%%%%%%%%%%%%%%%%%%%%%%%%%%%%% \noindent
{\bf (B) Limits of $x\gg v_1,\ v_2$($\l x$, $kx$ fixed)}\par
This limit leads the NMSSM to the MSSM without the Higgs singlet field as
discussed Ellis et al.[16].
In this limit with $A_\l, A_k\simeq O(v)$, the $X$ and $Y$  boson decouple
from other bosons, and then the matrix $M^2_{\rm Higgs}$ in eqs.(18)$\sim$(21)
reduces to the same $3\times 3$ matrix in eq.(22).
However,  masses of  $X$ and $Y$ are same order of other Higgs bosons in
contrast with the case (A).
Using the same orthogonal matrix in eq.(23), we get also the
similar matrix as the one in eq.(24) for the $A-S'_1-S'_2$ system as follows:
\begin{equation}
M'^2_{\rm Higgs} = \left[\matrix{ {2\bar\l A_\s\o\sin 2\b} & \bar\l \ \bar k
\sin\vp & 0 \cr & & \cr
\bar\l \ \bar k\sin\vp& (\bar g^2 \cos^2 2\b+\l^2\sin^2 2\b)v^2 &
(\l^2-\bar g^2)v^2\sin 2\b \cos 2\b \cr
& & \cr
0 & (\l^2-\bar g^2)v^2\sin 2\b \cos 2\b
&(\bar g^2-\l^2)v^2\sin^2 2\b+{2\bar\l A_\s\o\sin 2\b} \cr}\right ] \ ,
\end{equation}
\noindent
 where the definitions $\bar\l\equiv \l x$ and $\bar k\equiv k x$ are fixed
to be constants, while $\l$ and $k$ are order of $O(1/x)$ .
In contrast with the matrix of eq.(22), this matrix has not a hierarchical
structure in the considering limit since $\bar\l$ and $\bar k$ are finite
numbers.
Therefore, the submatrix of $S_1'-S_2'$ in eq.(28) are far from the diagonal
matrix in general.
Now, let us discuss the magnitude of  $CP$ violation for the special
case of  $\tan\b$.
\par
%%%%%%%%%%%%%%%%%%%%%%%%%%%%%%%%%%%%%%%%%%%%%%%%%%%%%%%%%%%%%
The first case is the one with $\tan\b=0$ and $\infty$.
Since $\sin 2\b=0$, the submatrix of the $S_1'-S_2'$ system is exactly
diagonal.
The scalar-pseodoscalar mixing is occured only in the $A-S_1'$ submatrix.
The mixing angle is given as follows:
\begin{equation}
\tan 2\th_{A S_1'}={2\bar\l\ \bar k\sin\vp
\o (\bar g^2 \cos^2 2\b+\l^2 \sin^2 2\b)v^2 -
{2\bar\l A_\s\o\sin 2\b}} \simeq -{\bar k\o A_\s} \sin\vp\sin 2\b \ .
\end{equation}
Thus, the scalar-pseudoscalar mixing vanishes in $\tan\b=0$ or $\infty$ limit
since it is proportional to $\sin 2\b$ even if $\sin\vp\simeq 1$.
Then, the $CP$ violation effect is expected generally to vanish.
However, we should pay attention to an exceptional case that the $CP$
violating effect depends on $\tan\b$ significantly.
We will discuss this case in analyses of the electric dipole moments of
the section 4.
\par
%%%%%%%%%%%%%%%%%%%%%%%%%%%%%%%%%%%%%%%%%%%%%%%%%%%%%%%%%%%%%%%
The second case is the one of $\tan\b=1$, which gives $\cos 2\b=0$.
In this case, the scalar-pseodoscalar mixing is also occured only in
the $A-S_1'$ submatrix since the $S_1'-S_2'$ submatrix is exactly diagonal.
Then, the $S_1'-S_1'$ component is $\l^2 v^2$ which is order of $O(v^4/x^2)$.
This hierarchical structure of the mass matrix gives strong constraint
for the mixing angle as discussed in the limitting case (A).
Applying the positivity condition of the  Higgs scalar mass in eq.(26) leads
\begin{equation}
\sin\vp \leq O({v\o x}) \ .
\end{equation}
\noindent Thus,  $CP$ violation also vanishes in the case of $\tan\b=1$.
\par
%%%%%%%%%%%%%%%%%%%%%%%%%%%%%%%%%%%%%%%%%%%%%%%%%%
In order to get the finite $CP$ violation, we should choose the region of
$\tan\b\not= 0$, $1$ and $\infty$. If we could adjust the parameter such as
\begin{equation}
2\bar\l A_\s \simeq \bar g^2 v^2\cos^2 2\b \sin 2\b \ ,
\end{equation}
\noindent by choosing the suitable $\tan\b$, the large scalar-pseudoscalar
mixing is expected.
However, since the radiative correction  $V_{\rm top}$ becomes significant
in this situation, we shall give the numerical analyses  in section 4.
\par
\vskip 0.3 cm
%%%%%%%%%%%%%%%%%%%%%%%%%%%%%%%%%%%%%%%%%%%%%%%%%%%%%%%%%%%%%%%
\noindent {\bf (C) Limits of $x\ll v_1,\ v_2$}\par
In the $x=0$ limit with $A_\l, A_k\simeq O(v)$, the submatrix
$M^{AY}_{S_1 S_2 X}$ is described as
\begin{equation}
M^{AY}_{S_1 S_2 X} = \left(\matrix{ 0 & 0 & 0 \cr
0 & 0 & -2k\l v_1v_2\sin\vp \cr}\right ) \ , \end{equation}
\noindent where only the (2-3) component remains to be finite, that is,
the scalar-pseudoscalar mixing exists only in the $X$ and $Y$ mixing.
Since  squares of $X$ and $Y$ boson masses are order of $O(v^3/x)$ and other
matrix elements are at most $O(v^2)$, these bosons  decouple from
other Higgs bosons except for the case of $\sin 2\b=0$.
Ellis et al.[16] have found that the large $\tan\b$ or
$\cot\b$($\sin 2 \b=0$) is not allowed by studying the constraint  that the
symmetry-breaking vacuum is a deeper minimum than the symmetry vacuum in the
case of $x \ll v_1,\ v_2$. So, we do not need to consider the case of
$\sin 2\b=0$.
\par
The submatrix of the $X-Y$ system is given as follows:
\begin{equation}
M^2_{\rm Higgs}= \left(\matrix{ {\l A_\l v^2\o 2 x}\sin 2\b & -k\l v^2
  \sin 2\b\sin\vp \cr & \cr
 -k\l v^2\sin 2\b\sin\vp &{\l A_\l v^2\o 2 x}\sin 2\b +2\l
 k v^2\sin 2\b\cos \vp }\right ) \ .
\end{equation}
\noindent
Then, the the mixing angle is given as follows:
\begin{equation}
\tan 2\th_{XY}=-{2k\l v^2\sin 2\b\sin\vp \o 2k\l v^2\sin 2\b\cos\vp }
=-\tan\vp \ ,
\end{equation}
\noindent
 so the maximal mixing of the scalar-pseudoscalar is realized in the case of
$\vp=\pm\p/2$.
The phenomena induced by $X$ and $Y$ Higgs bosons may show the large $CP$
violation.
However, the mass eigenvalues are infinite in the $x=0$ limit, so the $CP$
violation effect on the low energy phenomena would vanish.
\par
\vskip 1 cm
%%%%%%%%%%%%%%%%%%%%%%%%%%%%%%%%%%%%%%%%%%%%%%%%%%%%%%%%
\noindent{\bf 4.  Numerical Discussion of  Explicit $CP$ violation}
\par
In this section, we show the numerical examples to realize the  large $CP$
violation. Generally, the large $CP$ violation could be caused by
choosing $x$ to be $O(v)$.
However, then, the  Higgs boson spectroscopy  is very different from the MSSM
because  $S_1$, $S_2$ and $A$ bosons mix significantly with $X$ and $Y$ bosons.
In our interest, we present numerical study of the similar case to the MSSM
spectroscopy,  but the case with $CP$ violation.
This is just the limit in case (B).
\par
 In the previous section, we have neglected the radiatively induced potential
$V_{\rm top}$ for simplicity because the qualitative result is not changed
even if we include it.
Now, we should include the $V_{\rm top}$ term in our numerical analyses.
In the leading-log approximation, this potential contributes only to the
mass matrix element $M^{S_2}_{S_2}$ in eq.(20) as follows:
\begin{equation}
M^{S_2}_{S_2}=(\bar g^2+\Delta) v_2^2 +{\l v_1 x\o v_2} (A_\l+kx\cos\vp)\ ,
\end{equation}
\noindent
where
\begin{equation}
\Delta ={3 h_t^4\o 4\pi^2}\left [ \ln \left ( {M^2_{sq}\o m_t^2}
                           \right )+ p \right ] \ ,
\end{equation}
\noindent
where $p$ denotes non-logarithmic terms.
In the following calculations, we fix $\Delta=0.5$, which corresponds to
$M_{sq}=3 \rm{TeV}$  and $m_t=175 \G$ with $p=1$[14].
\par
%%%%%%%%%%%%%%%%%%%%%%%%%%%%%%%%%%%%%%%%%%%%%%%
\begin{center}
 \unitlength=0.7 cm
 \begin{picture}(2.5,2.5)
  \thicklines
  \put(0,0){\framebox(3,1){\bf Fig.1}}
 \end{picture}
\end{center}
\par
%%%%%%%%%%%%%%%%%%%%%%%%%%%%%%%%%%%%%%%%%%%%%%
In Fig.1, we display a plot of the experimentally allowed region in the
 $\cos\vp-\l$ plane for fixed values of the other parameters,
which are
\begin{equation}
 x=10 v, \quad k=0.1,\quad A_{\l}=v, \quad A_k=v, \quad  \tan\b=10 \ .
\end{equation}
  One experimental constraint is that the two Higgs bosons have not
been produced in the decay of a real $Z^0$[11].
The lower boundary(small $\l$)  in Fig.1 corresponds to
$m_{h_1}+m_{h_2}=m_{Z^0}$, where $m_{h_1}$ and  $m_{h_2}$ are two lightest
Higgs boson masses.
The other constraint is that a light Higgs boson has not been produced
in the $Z^0\A Z^{0*} h$ process, where $h$ is a physical Higgs boson.
If $h=\sum_{i=1}^{5}\a_i \Phi_i$, where $\a_i$ and $\Phi_i$ denote  mixing
factors and neutral Higgs boson fields $S_1, S_2, A, X, Y$, respectively,
the cross section for this process is approximately proportional to
 $|\a_1\cos\b+\a_2\sin\b|^2 m_h^{-1}$. The non-observation of this process
 gives the upper boundary(large $\l$) in Fig.1 by
$m_h\geq (60\G)|\a_1\cos\b+\a_2\sin\b|^2$[11].
 In addition, the pseudoscalar and scalar bosons should be heavier than
 $24\G$ and $44\G$, respectively[11].
This constraints are satisfied in the allowed region of Fig.1.
\par
In Fig.2, the allowed region of $\l$ is shown in the case of
$\tan\b=1\sim 100$ at $\cos\vp=0$.
Other parameters are fixed as given in eq.(37).
It is remarked that the allowed region vanishes  below $\tan\b\simeq 1.5$.
This result is consistent with the qualitaive discussion of (B) in section 3,
 in which  $\vp$ is constrained to be very small at  $\tan\b\simeq 1$,
 and $\vp\simeq \pi/2$ is allowed at $\tan\b=\infty$.
In both results of Figs. 1 and 2, we fix $k=0.1$, which  gives the
most wide allowed area of $\l$.
As far as we take $k=0.03 \sim 0.2$, the allowed region is obtained.
\par
%%%%%%%%%%%%%%%%%%%%%%%%%%%%%%%%%%%%%%%%%%%%%%%
\begin{center}
 \unitlength=0.7 cm
 \begin{picture}(2.5,2.5)
  \thicklines
  \put(0,0){\framebox(3,1){\bf Fig.2}}
 \end{picture}
\end{center}
\par
%%%%%%%%%%%%%%%%%%%%%%%%%%%%%%%%%%%%%%%%%%%%%%
The electric dipole moment(EDM) of electron or neutron is very important
 quantities to constrain the phase $\vp$.  In our scheme, the EDM of electron
is calculated  in the two-loop level  as shown by Barr and Zee[18].
The neutron EDM is also predicted in two-loop level.
Both  three gluon operator proposed by Weinberg[19] and  quark-gluon operator
  by Gunion and Wyler[20] are taken into account  in our calculation.
Since the estimation of the hadronic matrix elements is model-dependent,
 the ambiguity with a few factors should be  taken into consideration in the
prediction of the neutron EDM.
Here,  we use the model proposed by Chemtob[21,22].
The recent experimental upper limit of the electron EDM is
$4\times 10^{-27} {\rm e\cdot cm}$[23] and that of the neutron EDM
 is $11\times 10^{-26} {\rm e\cdot cm}$[11].
It should be remarked that the Barr-Zee operator and the quark-gluon operator
 are  exceptional $CP$ violating operators as discuussed in (B) of section 3.
 Since these operators have a term which is proportional to  $\tan^2\b$,
this term contributes to the EDM  significantly at $\tan\b\gg 1$
 even if the scalar-pseusoscalar mixing is very small.
 In fact, we   find the large predicted EDM at $\tan\b=10$
in Figs. 3 and 4.
In these figures, we give the numerical predictions of the electron EDM
and the neutron EDM  in the allowed region of $\l$ in Fig.1.
The upper(lower) boundary of the predictions corresponds to the upper(lower)
 one of $\l$ in Fig.1.
Those predictions lie around experimental upper limits except for
 the region of  $\cos\vp\simeq \pm 1$. If the  small $\l$, $O(0.01)$, is taken,
 our predictions are below the experimental limits even if
 the phase $\vp$ is a maximal one $\pi/2$.
We expect both electron EDM and  neutron EDM will be observed around
$10^{-27}\sim 10^{-26}{\rm e\cdot cm}$ in the near future.
\par
%%%%%%%%%%%%%%%%%%%%%%%%%%%%%%%%%%%%%%%%%%%%%%%
\begin{center}
 \unitlength=0.7 cm
 \begin{picture}(2.5,2.5)
  \thicklines
  \put(-1,0){\framebox(5,1){\bf Figs.3 and 4}}
 \end{picture}
\end{center}
\par
%%%%%%%%%%%%%%%%%%%%%%%%%%%%%%%%%%%%%%%%%%%%%%
 Let us discuss about the mass of  the lightest Higgs boson in our scheme.
 What is the $CP$ violating effect on it ?  The top-loop effect
seems to be significant in the lightest Higgs boson mass
 in the case without $CP$ violation[17].
We show the mass of the lightest Higgs boson versus $\l$
 for fixed $\cos\vp=1,\  0.5,\  0,\  -0.5$ in Fig.5.
 In the region of $\l=0.01\sim 0.1$, the $CP$ violating effect
 reduces the magnitude of the mass in the order of
 $10\sim 20\G$.
The qualitative result is not so changed if we take $M_{\rm sq}\ll 3$TeV.
\par
%%%%%%%%%%%%%%%%%%%%%%%%%%%%%%%%%%%%%%%%%%%%%%%
\begin{center}
 \unitlength=0.7 cm
 \begin{picture}(2.5,2.5)
  \thicklines
  \put(0,0){\framebox(3,1){\bf Fig.5}}
 \end{picture}
\end{center}
\par
%%%%%%%%%%%%%%%%%%%%%%%%%%%%%%%%%%%%%%%%%%%%%%
%\vskip 0.5 cm
\newpage
\noindent{\bf 5.  Summary}
\par
 We have studied the explicit $CP$ violation of the Higgs sector in the
MSSM with a gauge singlet Higgs field.
The magnitude of $CP$ violation
 is discussed in the limiting cases of $x\gg v_1,\ v_2$ and $x\ll v_1,\ v_2$.
We have shown that the large $CP$ violation is realized in the region of
$\tan\b\geq 1.5$ for the case of $x\gg v_1,\ v_2$ with the fixed values of
$\l x$ and $kx$.
In other cases, the explicit $CP$ violation is minor for the Higgs sector.
Since $CP$ violation in the Higgs sector does not ocuur in the MSSM
 without a gauge singlet Higgs field,
$CP$ violation is an important signal of the existence of
the gauge singlet Higgs field.
 In the present case of the Higgs sector, the predictions of
the electron EDM and the neutron EDM lie around the experimental upper limits.
 Our results suggest that these EDM's will be observed in the near future
if $CP$ is explicitly violated through the Higgs sector in the NMSSM.
Furthermore, we have found that the large $CP$ violation effect reduces
the magnitude of the lightest Higgs boson mass in the order of
 a few ten GeV.
Thus, the explicit $CP$ violation due to the gauge singlet Higgs boson
 will give us interesting phenomena in the forthcoming experiments.
\par
\vskip 0.5 cm
%%%%%%%%%%%%%%%%%%%%%%%%%%%%%%%%%%%%%%%%%%%%%%%%%%%%%%%%
%%%%%%%%%%%%%%%%%%%%%%%%%%%%%%%%%%%%%%%%%%%%%%%%%%%%%%%%
 We would like to thank Professors
 T. Hayashi, Y. Koide and  S. Wakaizumi and Mr. N. Haba
for helpful discussions.
 This research is supported by the Grant-in-Aid for Scientific
Research, Ministry of Education, Science and Culture,
Japan(No.06220101 and No.06640386).
\newpage
%%%%%%%%%%%%%%%%%%%%%%%%%%%%%%%%%%%%%%%%%%%%%%%%%%%%%%%%%%
%%%%%%%%%%%%%%%%%%%%%%%%%%%%%%%%%%%%%%%%%%%%%%%%%%%%
%%%%%%%%%%%%%%%%%%%%%%%%%%%%%%%%%%%%%%%%%%%%%%%%%%%%%%%
\newpage
\centerline{{\large \bf References}}
\vskip 0.5 cm
\noindent
[1] M.Kobayashi and T.Maskawa, Prog. Theor. Phys.
{\bf 49}, 652(1973).\par
\noindent
[2] For a text of Higgs physics see J.F. Gunion, H.E. Haber, G.L.Kane   and S.
Dawson,\par
{\it "Higgs Hunter's Guide"}, Addison-Wesley, Reading, MA(1990). \par \noindent
[3] T.D. Lee, Phys. Rev. {\bf D8}, 1226(1973).\par \noindent
[4] G.C. Branco and M.N. Rebelo, Phys. Lett.
{\bf 160B}, 117(1985).
\par\noindent
[5] S. Weinberg, Phys. Rev. {\bf D42}, 860(1990).\par \noindent
[6] U. Amaldi, W. de Boer and H. F\"urstenau,
  Phys. Lett. {\bf 260B}, 447(1991).
\par\noindent
[7] H.P. Nilles, Phys. Rep. {\bf 110}, 1(1984);\par H.E. Haber and
 G.L. Kane, Phys. Rep. {\bf 117}, 75(1985);\par
 A.B. Lahanas and D.V. Nanopoulos, Phys. Rep. {\bf 145}, 1(1987);\par
 J.F. Gunion and H.E. Haber, Nucl. Phys. {\bf B272}, 1(1986);
 {\bf B278}, 449(1986); \par
 {\bf B307}, 445(1988).\par
 \noindent
[8]  N.Maekawa, Phys. Lett. {\bf B282}, 387(1992).  \par
  \noindent
[9] A. Pomarol, Phys. Lett. {\bf B287}, 331(1992).
    \par \noindent
[10] H. Georgi and S.L. Glashow, Phys. Rev. {\bf D6}, 2977(1972); \par
   S. Coleman and E. Weinberg,  Phys. Rev. {\bf D7}, 1888(1973); \par
 H. Georgi and A. Pais,  Phys. Rev. {\bf D10}, 1246(1974); \
  {\bf D16}, 3520(1977). \par
\noindent
[11] Particle Data Group, K. Hikasa et al.,  Phys. Rev.
   {\bf D49}, S1(1994).\par
 \noindent
[12] S.M. Barr and A. Masiero, Phys. Rev. {\bf D38}, 366(1988); \par
  A. Pomarol, Phys. Rev. {\bf D47}, 273(1992); \par
  J.R. Espinosa, J.M. Moreno and M. Quiros,
  Phys. Lett. {\bf B319}, 505(1993); \par
  S.M. Barr and G. Segre, Phys. Rev. {\bf D48}, 302(1993); \par
 A. Pomarol, Phys. Rev. {\bf D47}, 273(1993).
    \par \noindent
[13] J.C. Rom\~ao, Phys. Lett. {\bf B173}, 309(1986). \par
\noindent
[14] K.S. Babu and S.M. Barr,  Phys. Rev. {\bf D49}, 2156(1994). \par
\noindent
[15] P. Fayet, Nucl. Phys. {\bf B90}, 104(1975);\par
 R.K. Kaul and P. Majumdar, Nucl. Phys. {\bf B199}, 36(1982);
   \par
 R. Barbieri, S. Ferrara and C.A. S\'avoy,
  Phys. Lett. {\bf 119B}, 343(1982);
  \par
  H.P. Nilles, M. Srednicki and D. Wyler,
  Phys. Lett. {\bf 120B}, 346(1983);\par
  J.M. Fr\`ere, D.R.T. Jones and S. Raby,
  Nucl. Phys. {\bf B222}, 11(1983);\par
  J.P. Derendinger and C.A. Savoy, Nucl. Phys. {\bf B237}, 307(1984).
\par \noindent
[16] J. Ellis, J.F. Gunion, H.E. Haber, L. Roszkowski and
F. Zwirner, Phys. Rev. \par
{\bf D39}, 844(1989).
\par
\noindent
[17] Y. Okada, M. Yamaguchi, and T. Yanagida, Prog. Theor. Phys. {\bf 85},
 1(1991); \par
 J. Ellis, G. Ridolfi and F. Zwirner,  Phys. Lett. {\bf 257B}, 83(1991);
\par
H. Haber and R. Hempfling, Phys. Rev. Lett. {\bf 66}, 1815(1991); \par
R. Barbieri, M. Frigeni and F. Caravaglios,
  Phys. Lett. {\bf 258B}, 167(1991); \par
A. Yamada,   Phys. Lett. {\bf 263B}, 233(1991); \par
P. Bin\'etruy and C.A. Savoy,  Phys. Lett. {\bf 277B}, 453(1992).
\par\noindent
[18] S.M. Barr and A. Zee, Phys. Rev. Lett. {\bf 65}, 21(1990);\par
   S.M. Barr, Phys. Rev. Lett. {\bf 68}, 1822(1992);
            Phys. Rev. {\bf D47}, 2025(1993);\par
  D. Chang, T.W. Kephart, W-Y. Keung and T.C. Yuan,
                    Phys. Rev. Lett. {\bf 68},\par
          439(1992).\par
\noindent
[19] S. Weinberg, Phys. Rev. Lett. {\bf 63}, 2333(1989).\par
\noindent
[20] J.F. Gunion and D. Wyler, Phys. Letts. {\bf 248B}, 170(1990).\par
\noindent
[21] M. Chemtob, Phys. Rev. {\bf D45}, 1649(1992).\par
\noindent
[22] T.Hayashi, Y.Koide, M.Matsuda and M.Tanimoto,
     Prog. Theor. Phys. {\bf 91}, \par
                                  915(1994).\par
 \noindent
[23]  E.D. Commins, S.B. Ross, D. DeMille and B.C. Regan,
  Phys. Rev. {\bf A50},\par
 2960(1994).\par

\newpage

\centerline{\large \bf Figure Captions}\par
\vskip 0.6 cm
\noindent
{\bf Fig.1}\par
 The allowed region in the
 $\cos\vp-\l$ plane for  $x=10 v$, $k=0.1$ $A_{\l}=v$, $A_k=v$ and
$\tan\b=10$.  The lower boundary  corresponds  to  $m_{h_1}+m_{h_2}=m_{Z^0}$
  and   the upper boundary is given by
 the nonobservation of $Z^0\A Z^{0*} h$.
\par
\vskip 0.5 cm
\noindent
{\bf Fig.2}\par
 The allowed region in the  $\tan\b-\l$ plane at $\cos\vp=0$.
  The notations are same as in Fig.1
\par
\vskip 0.5 cm
\noindent
{\bf Fig.3}\par
The predicted electron EDM in the allowed region  in Fig.1.
  The lower(upper) boundary corresponds to the lower(upper) one in Fig.1.
The doted-line denotes the experimental upper-limit.
\par
\vskip 0.5 cm
\noindent
{\bf Fig.4}\par
The predicted  neutron EDM in the allowed region in Fig.1.
  The lower(upper) boundary corresponds to the lower(upper) one in Fig.1.
The doted-line denotes the experimental upper-limit.
\par
\vskip 0.5 cm
\noindent
{\bf Fig.5}\par
The predicted lightest Higgs boson mass versus $\l$
 for $\cos\vp=1,\  0.5,\  0,\  -0.5$ in the case of $\Delta=0.5$.

\end{document}